\documentclass{aa}
\usepackage{graphicx}
\usepackage{txfonts}

\begin{document}

\title{On the hydrogen neutral outflowing disks of B[e] supergiants}

\author{M. Kraus\inst{1}, M. Borges Fernandes\inst{2}
  \and
   F.X. de Ara\'ujo\inst{3}
  }

\offprints{M. Kraus, \\ \email{kraus@sunstel.asu.cas.cz}}

\institute{Astronomical Institute AV \v{C}R, Fri\v{c}ova 298, 251\,65 Ond\v{r}ejov, Czech Republic
 \and
    Observat\'orio do Valongo (UFRJ), Ladeira do Pedro Antonio 43, Rio de Janeiro, 20080-090, Brazil
 \and
   Observat\'orio Nacional, Rua General Jos\'e Cristino 77, 20921-400 S\~ao Cristov\~ao, Rio de Janeiro, Brazil\\
 \email{kraus@sunstel.asu.cas.cz, borges@ov.ufrj.br, araujo@on.br}
      }

\date{Received; accepted}

\abstract
{B[e] supergiants are known to possess geometrically thick dusty disks.
Disk-forming wind models in the literature have, however, been found to be 
insufficient in reproducing the observed dust emission. This problem arises 
due to the severe assumption that, as for classical Be stars, the near-infrared 
excess emission originates in the disk. Modeling of the free-free and 
free-bound emission therefore results in an upper limit for the disk mass loss 
rate as well as for the disk opacity. Dust condensation in the disk can thus 
severely be hampered.}
{In order to overcome the dust formation problem, and based on our 
high-resolution optical spectroscopy and model results, we propose a revised 
scenario for the non-spherical winds of B[e] supergiants: a normal B-type 
line-driven polar wind and an outflowing disk-forming wind that is 
neutral in hydrogen at, or very close to the stellar surface.}
{We concentrate on the pole-on seen LMC B[e] supergiant R\,126 and calculate the
line luminosities of the optical [O{\sc i}] emission lines and their emergent 
line profiles with an outflowing disk scenario. In addition, we compute the 
free-free and free-bound emission from a line-driven polar wind and model the 
spectral energy distribution in the optical and near-infrared.}
{Good fits to the [O{\sc i}] line luminosities are achieved for an outflowing
disk that is neutral in hydrogen right from the stellar surface. Neutral thereby
means that hydrogen is ionized by less than 0.1\%. Consequently, the free-free 
and free-bound emission cannot (dominantly) arise from the disk and cannot limit
the disk mass loss rate. The hydrogen neutral outflowing disk scenario therefore
provides an ideal environment for efficient dust formation. The spectral energy 
distribution in the optical and near-infrared range can be well fitted with the 
stellar continuum plus free-free and free-bound emission from the polar 
line-driven wind. Our modeling further delivers minimum values for $\dot{M}_{\rm
disk} \ga 2.5\times 10^{-5}$\,M$_{\odot}$yr$^{-1}$ and for the density contrast between equatorial and polar wind of $\sim 10$.}
{}



\maketitle

\section{Introduction}

The spectra of B[e] supergiants show the so-called hybrid character which is 
defined by the co-existence of a line-driven polar wind and a high density but 
low velocity equatorial wind (Zickgraf et al. \cite{Zickgraf}). The latter is 
assumed to form a disk-like structure. This disk is the location of the
low-ionized metals, of the sometimes observed CO and TiO molecular emission 
bands (McGregor et al. \cite{McGregor88}, \cite{McGregor89}; Zickgraf et al. 
\cite{Zick_S18}) as well as of the hot dust, pronounced in the mid-IR 
excess emission (e.g. Zickgraf \cite{Zick92}).
The disk formation mechanism, as well as the disk structure, are however still
unsolved problems. 

Porter (\cite{John}) has studied the possible nature of these disks by modeling 
their spectral energy distribution (SED). He found that neither an outflowing 
disk-forming wind nor a Keplerian viscous disk in their simplest form can 
easily account for the observed free-free and dust emission self-consistently.  

\begin{figure}[t!]
\resizebox{\hsize}{!}{\includegraphics{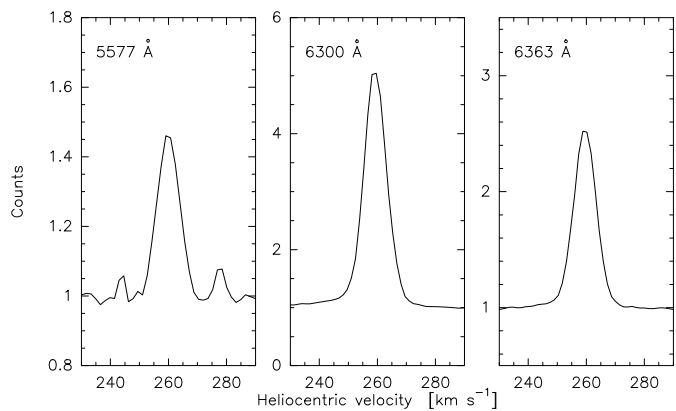}}
\caption{The [O{\sc i}] lines in the FEROS spectrum of R\,126. They all show a
redshift of about 260\,km\,s$^{-1}$ caused by the radial velocity of the LMC.}
\label{oi_r126}
\end{figure}

The optical spectra of B[e] supergiants exhibit strong [O{\sc i}] emission
(see e.g. Kraus \& Borges Fernandes \cite{KrausBorges}; Kraus et al. 
\cite{Vlieland}; and Fig.\,\ref{oi_r126}). Besides other typical features like 
strong Balmer emission, Fe{\sc ii} and [Fe{\sc ii}] emission, and mid-IR excess 
emission, the [O{\sc i}] emission lines are one of the main characteristics of 
stars with the B[e] phenomenon (Lamers et al. \cite{Lamers98}).
The fact that O{\sc i} and H{\sc i} have about equal ionization 
potentials requests that the [O{\sc i}] emission region must be neutral in 
hydrogen. Its most plausible location is therefore the high-density disk.

Test calculations for an outflowing disk scenario, that were aimed to reproduce 
the observed strong [O{\sc i}] line luminosities, have been performed for some 
Magellanic Cloud B[e] supergiants (Kraus \& Borges Fernandes \cite{KrausBorges};
Kraus et al. \cite{Vlieland}). These computations revealed that, in order to 
keep the disk mass loss rates at reasonable values, the disk material 
must be neutral in hydrogen either at or at least very close to the stellar 
surface, resulting in a {\it hydrogen neutral disk}. 

In this paper we present model results for the Large Magellanic Cloud (LMC) B[e]
supergiant R\,126. We use a simple outflowing disk scenario and request that
the disk is neutral in hydrogen right from the stellar surface. 
We show that with such a model we can explain observations like the strong 
[O{\sc i}] line luminosities. In addition, this scenario provides a way out
of the dust formation problem mentioned by Porter (\cite{John}). 
The paper is structured as follows: 
in Sect.\,\ref{target} we give details on the test star and on our observations.
Sect.\,\ref{motivation} introduces the problem of the nature of the B[e] 
supergiants disks. We emphasize the need of a revised disk model and propose
the hydrogen neutral disk as a reasonable scenario.
In Sect.\,\ref{neutral} we present our hydrogen neutral disk model and 
calculate the emerging [O{\sc i}] line luminosities as well as their line
profiles. In Sect.\,\ref{polar} we fit the SED in the optical and near-IR range
by calculating the free-free and free-bound emission from the polar
wind. In addition, we determine the density ratio between equatorial and polar 
wind components and show that our model results in a lower limit. The 
validity of our assumptions is discussed in Sect.\,\ref{discussion}, and 
the conclusions are summarized in Sect.\,\ref{conclusions}.

\section{The LMC B[e] supergiant R\,126}\label{target}
                                                                                
\begin{figure}[t!]
\resizebox{\hsize}{!}{\includegraphics{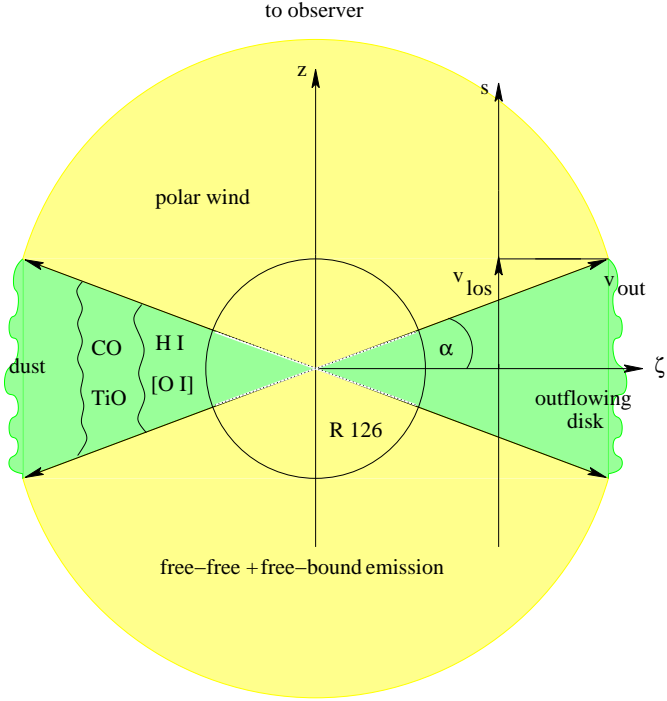}}
\caption{Sketch of the pole-on seen disk around R\,126 (not to scale). The disk
(dark grey shaded) is assumed to be neutral in hydrogen right from the stellar 
surface, giving raise to [O{\sc i}] emission. Further out, molecules and dust 
will form. The disk material moves outwards with $v_{\rm out}$, resulting in an 
observable line-of-sight velocity ($v_{\rm los}$). Free-free and free-bound 
emission is restricted to the (polar) wind (light grey shaded). The system is 
axially symmetric (around $z$-axis) and symmetric with respect to the disk 
mid-plane.
}
\label{sketch}
\end{figure}

The LMC B[e] supergiant R\,126 (HD\,37974, LHA 120-S 127) is supposed to have a
pole-on seen disk. This orientation has the following advantages (see 
Fig.\,\ref{sketch}): (i) it guarantees that we see all of the [O{\sc i}] 
emission and none is hidden by the star or absorbed by the dusty disk  
because the [O{\sc i}] emission originates from regions close to the star where 
no dust has yet formed, and (ii) it simplifies the analysis due to the high 
degree of symmetry. R\,126 has also been studied by Porter (\cite{John}) so 
that we can directly compare our results. The stellar parameters of this object 
have been derived by Zickgraf et al. (\cite{Zickgraf}) who found: $T_{\rm eff} 
= 22\,500$\,K, $L_* = 1.2\times 10^{6}$\,L$_{\odot}$, $R_{*} = 72$\,R$_{\odot}$,
and an interstellar extinction value of $E(B-V) = 0.25$.
These values will be used for our calculations throughout the paper.

We obtained high- and low-resolution optical spectra of R\,126 at the ESO
1.52-m telescope (agreement ESO/ON-MCT) in La Silla (Chile) using the Fiber-fed
Extended Range Optical Spectrograph (FEROS) and the Boller \& Chivens
spectrograph, respectively. FEROS is a bench-mounted Echelle spectrograph with
fibers, that cover a sky area of 2 arcsec, located at the Cassegrain focus with
a wavelength coverage from 3600\,\AA \ to 9200\,\AA \ and a spectral resolution
of R = 55\,000 (in the region around 6000 \AA). It has a complete automatic
on-line reduction, which we adopted. The spectra were taken on December 18,
1999, with an exposure time of 4500 seconds. The S/N ratio in the 5500\,\AA \
region is approximately 100. 

The low-resolution Boller \& Chivens (B\&C) spectrum was taken on October 31, 
1999, with an exposure time of 900 seconds and a slit width of 4 arcsec. The 
instrumental setup employed provides a resolution of $\sim 4.6$\,\AA \ in the 
range of 3800-8700\,\AA. In the 5500\,\AA \ continuum region, the S/N ratio is 
aproximately 200. The B\&C spectrum was reduced using standard IRAF 
tasks, such as bias subtraction, flat-field normalization, and wavelength 
calibration. We have done absolute flux calibration using spectrophotometric 
standards from Hamuy et al. (\cite{Hamuy}).

Equivalent widths and line intensities in both linearized spectra have been 
measured using the IRAF task that computes the line area above the adopted 
continuum. Uncertainties in our measurements come mainly from the position of 
the underlying continuum and we estimate the errors to be about 20\,\%. The 
emerging [O\,{\sc i}] lines resolved within the FEROS spectrum are shown in 
Fig.\,\ref{oi_r126}.

\section{On the nature of B[e] supergiant stars' disks}\label{motivation}

The nature of the B[e] supergiant stars' disks is a long-standing problem.
Recently, Porter (\cite{John}) investigated the possibility of dust formation 
in the disk of the B[e] supergiant R\,126 for two different approaches: an 
outflowing disk-forming wind, and a Keplerian viscous disk. He found that both 
models failed in reproducing the observed dust and free-free emission 
self-consistently. And he suggested a way out of this by allowing for 
substantial alteration especially of the exponent of the radial density 
distribution from $\rho\sim r^{-2}$ to a considerably flatter one of 
$r^{-1.7}$. The advantage of a flatter density profile is that at every 
location in the disk a higher density (and hence opacity) is maintained, 
allowing for more efficient dust condensation and therefore for an enhanced 
dust emission over the disk. Such a modification finally resulted in a good
fit to the SED of R\,126 (see Fig.\,6 of Porter \cite{John}), but for the
price that the disk density distribution now no longer corresponds to
a constant velocity outflow (which requests $\rho\sim r^{-2}$). 
A possible interpretation of
such a flatter density profile might be that the disk slows down with distance, 
a scenario that lacks any observational evidence.

To test Porter's modified outflowing disk scenario, we took the 
parameters of his best-fit model and calculated the emerging [O{\sc i}] line 
luminosities. These turned out to be at least a {\it factor of 50 lower} than 
the values we observed. This means that we need a much higher density within 
the [O{\sc i}] line forming region than can be provided by Porter's disk model.
 
The most severe limitation in his model calculations was the assumption 
that, as in the case of classical Be stars, the free-free emission arises in 
the high density disk, while contributions from the polar wind are negligible. 
Fitting the near-IR part of the SED with free-free emission from the outflowing 
disk therefore determines the disk mass loss rate. Consequently, this value is 
an {\it upper limit} for the disk density and hence opacity. Dust can only 
condensate at distances with reasonable values of temperature and opacity. An
upper limit for the disk density, defined by the free-free emission, can 
therefore severely hamper efficient dust formation in the disk, which turned 
out to be the major reason why Porter (\cite{John}) could not fit the observed 
dust emission of R\,126 with the original outflowing disk scenario.

The detection of strong [O{\sc i}] line emission in the spectra of e.g. B[e] 
supergiants requests that there must be a rather high density disk 
region which is neutral in hydrogen ($T \la 10000$\,K), but still hot enough 
($6000\la T\,[{\rm K}] \la 8000$) for effective excitation of the levels in
O{\sc i}. First test calculations that were aimed to reproduce the [O{\sc i}] 
line luminosities with an outflowing disk scenario have been performed by Kraus 
\& Borges Fernandes (\cite{KrausBorges}) and Kraus et al. (\cite{Vlieland}). 
These computations showed that, in order to keep the disk mass loss rates at 
reasonable values, the disks must be neutral in hydrogen either at,
or at least very close to the stellar surface. 

The existence of hydrogen neutral disks around B[e] supergiants is additionally
supported by recent ionization structure calculations in non-spherically
symmetric winds. These computations have shown that the wind material in the
equatorial plane can indeed recombine at, or close to the stellar surface if
either the equatorial mass flux is enhanced compared to the polar one
(Kraus \& Lamers \cite{KL03}), or
even decreased in combination with a reduction in surface temperature due to
rapid rotation of the star (Kraus \cite{Kraus06}; Kraus \& Lamers \cite{KL06}).

Having a hydrogen neutral disk around a luminous object is therefore not
as unrealistic as it might seem at first glance. In fact, it has several 
striking advantages: if the disk is predominantly neutral,
the free-free and free-bound emission can no longer be generated
in the disk, but must (mainly) originate within the polar wind. 
The disk mass loss rate is therefore {\it not strictly linked} to the 
free-free emission causing the near-IR excess.
Hence, the disk mass loss rate might be (much) higher than the upper
limit used by Porter (\cite{John}).
A higher disk density is certainly needed to reproduce the observed 
[O{\sc i}] line luminosities. And finally, a higher disk density will increase 
the disk opacity and will allow for 
more efficient dust condensation in the disk, needed to reproduce the observed
IR excess.

In the following we will investigate and test the suggested model scenario of a 
hydrogen neutral disk around a luminous object by analysing in detail the 
observed [O{\sc i}] lines from the B[e] supergiant R\,126.

\section{The (hydrogen) neutral disk}\label{neutral}

\subsection{The disk geometry}\label{geometry}

The disk is assumed to be wedge-shaped (see sketch in Fig.\,\ref{sketch})
and we use the same value of $\alpha = 10\degr$ for the constant disk half 
opening angle as Porter (\cite{John}). 
For simplicity we assume that the disk has a constant outflow velocity 
within the [O{\sc i}] emitting region. The influence of a velocity distribution
on the final results is discussed in Sect.\,\ref{discussion}.

In a disk-forming wind of constant outflow velocity, 
the hydrogen particle density distribution at all latitudes 
within the disk can be written in the form
\begin{equation}
n_{\rm H}(r) = n_{\rm H}(R_{\rm in}) \frac{R_{\rm in}^{2}}{r^2}\,. 
\label{densdistr}
\end{equation}
The parameter $n_{\rm H}(R_{\rm in})$, given by
\begin{equation}
n_{\rm H}(R_{\rm in}) = \frac{1}{\mu m_{\rm H}} \left(\frac{R_*}{R_{\rm in}}\right)^{2}\,\frac{F_{\rm m, disk}}{v_{\rm out}}  
\label{densparameter}
\end{equation} 
with the disk mass flux, $F_{\rm m, disk}$, and the (constant) disk 
outflow velocity, $v_{\rm out}$ (see Fig.\,\ref{sketch}),
describes the density at the inner radius, $R_{\rm in}$, of the 
[O{\sc i}] emitting region. This inner radius corresponds to the hydrogen 
recombination distance in the disk and is in principle a free 
parameter in our calculations, but here we use $R_{\rm in} = R_*$ since this 
defines a lower limit to the disk density needed to reproduce the [O{\sc i}]
line luminosities. The validity of our assumption that the disk is
neutral in hydrogen right from the stellar surface is discussed in 
Sect.\,\ref{discussion}.
The oxygen and electron density distributions are parameterized in terms
of the hydrogen density, i.e.
\begin{equation}
n_{\rm O}(r) = q_{\rm O} n_{\rm H}(r) \qquad {\rm and} \qquad 
n_{\rm e}(r) = q_{\rm e} n_{\rm H}(r)
\end{equation}
The parameters $q_{\rm O}$ and $q_{\rm e}$ describe the relative abundances of 
oxygen and electrons with respect to hydrogen. For $q_{\rm O}$ we use a typical 
LMC abundance value of 1/3 solar, with a solar oxygen abundance of 
$q_{\rm O, solar} = 6.76\times 10^{-4}$ (Grevesse \& Sauval \cite{Grevesse}). 
Since we do not know the ionization fraction of the disk a priori, we keep
$q_{\rm e}$ as a free parameter.

\subsection{The [O{\sc i}] lines and the disk ionization}

\begin{figure}[t!]
\resizebox{\hsize}{!}{\includegraphics{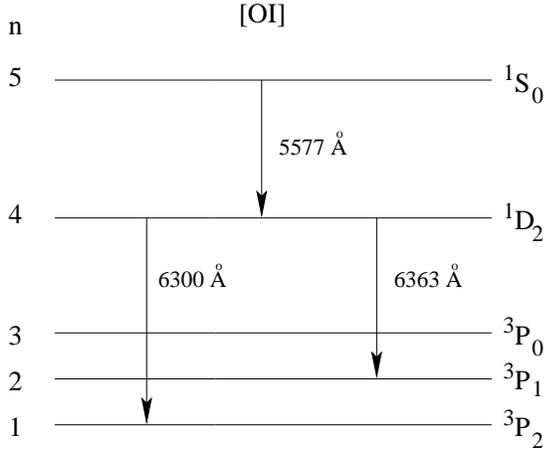}}
\caption{Sketch of the five lowest energy levels within O{\sc i} from which the
forbidden lines arise. The transitions of our interest are indicated.}
\label{levels}
\end{figure}

The forbidden emission lines arise from the five lowest energy levels of the
O{\sc i} atom that are excited via collisions with free electrons. These levels 
are shown schematically in Fig.\,\ref{levels}. For simplicity, we labeled them 
from 1 to 5 corresponding to their increasing energy. The three lines resolved 
in our high-resolution optical spectra are indicated; they belong to the 
transitions $5\longrightarrow 4$ (5577\,\AA), $4\longrightarrow 1$ (6300\,\AA), 
and $4\longrightarrow 2$ (6363\,\AA). 

While the lines 6300\,\AA~and 6363\,\AA~arise from the same upper level, which
means that their line ratio is independent of the density and temperature but 
is determined purely by atomic parameters, the situation is different for  
line ratios with the 5577\,\AA~line. This line arises from the highest energy 
level in our 5-level atom (Fig.\,\ref{levels}). It is therefore obvious that
the 5577\,\AA~line will only become strong if level 5 is populated efficiently, 
either due to a huge number of available free electrons or by electrons with 
rather high energy. Line ratios with the 5577\,\AA~line are thus sensitive
tracers for the temperature and ionization fraction in the disk.
In our spectra, the 6300\,\AA~line is the one with highest luminosity. We will
therefore use the 6300\,\AA/5577\,\AA~line luminosity ratio to investigate 
density and temperature in the disk around R\,126.

Forbidden lines are optically thin. The emissivity at any location $r$ in the
disk of a transition $n\rightarrow m$ is given by $j_{\rm n,m}(r) = h \nu_{\rm
n,m} n_{\rm n}(r) A_{\rm n,m}$ where $n_{\rm n}(r)$ is the level population of
the upper level, n, at this location. The level populations are calculated by
solving the statistical equilibrium equations in our 5-level atom. Collision
parameters are taken from Mendoza (\cite{Mendoza}) and atomic parameters from
Wiese et al. (\cite{Wiese}) and from Kafatos \& Lynch (\cite{Kafatos}). 
The line luminosity finally follows from integration of the emissivity over the
emitting disk volume.

Due to the nearly equal ionization potentials of O and H, the [O{\sc i}]
emission must arise in those parts of the disk in which hydrogen is 
predominantly neutral. This does not mean that hydrogen must be 100\% neutral,
since effects like collisional ionization that might take place in the high 
density disk could keep the hydrogen material ionized at a (very) low level, 
providing free electrons to collisionally excite O{\sc i}. Further sources 
of free electrons in the disk are elements like Mg, Si, Fe, Al, etc. with 
ionization potentials (much) lower than 13.6\,eV. These metals will remain 
ionized at a certain fraction even in a completely hydrogen neutral environment.
The ionization fraction, $q_{\rm e}$, in the disk is therefore the sum of
all free electrons provided by hydrogen and the metals, i.e. $q_{\rm e} = 
q_{{\rm H}^{+}} + q_{{\rm Metals}^{+}}$.

To get a handle on the possible individual contributions to the total number 
of free electrons in the disk, we make a first guess.
If all possible elements with ionization potential lower than 13.6\,eV would 
be fully singly ionized (including also C with $\chi = 11.26$\,eV 
and Cl with $\chi = 12.97$\,eV), then a maximum number of free electrons
of $q_{{\rm Metals}^{+}} \le 1.57\times 10^{-4}$ would be provided by the 
metals in the disk. This number accounts for an LMC metallicity of 1/3 solar.
On the other hand, if only 1\,\% of hydrogen in the disk remains ionized, the 
ionization fraction would be $q_{{\rm H}^{+}} = 10^{-2}$ which is at least a 
factor 100 higher than what can be expected from the metals. 
Our assumption of a hydrogen neutral disk around R\,126 therefore requests
that the total disk ionization fraction should be $q_{\rm e} < 10^{-2}$.  

\begin{figure}[t!]
\resizebox{\hsize}{!}{\includegraphics{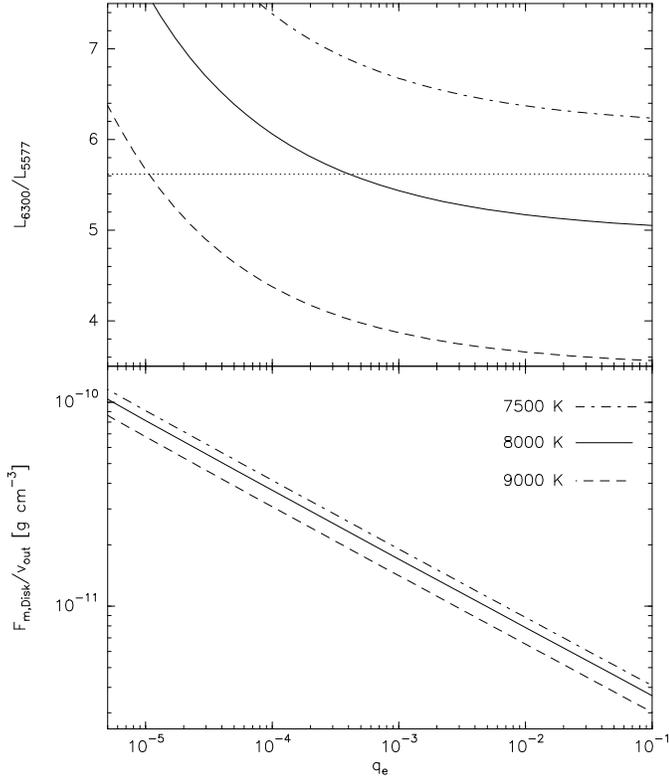}}
\caption{Upper panel: Variation of the line ratio with disk ionization fraction.
The three curves are for the different temperatures as indicated. The dotted
line represents the observed line ratio from R\,126. It indicates a disk
temperature in the range 9000\,K to about 8000\,K with corresponding ionization
fractions of $10^{-5}$ and $4\times 10^{-4}$, respectively. Lower panel:
Density parameter of the disk found from fitting the observed 6300\,\AA~line
luminosity over a huge range of ionization fractions. For details see text.}
\label{ionfrac}
\end{figure}

To derive the ionization fraction in the disk of R\,126 we calculated 
the line luminosities and the 6300\,\AA/5577\,\AA~line ratio for a large range
of $q_{\rm e}$ values and for different temperatures. These calculations were
done in several steps: for a given ionization fraction and temperature we first 
varied the density parameter, $F_{\rm m, disk}/v_{\rm out}$ (see 
Eq.\,\ref{densparameter}), in order to reproduce the
observed line luminosity of the 6300\,\AA~line. This
density parameter is plotted as a function of $q_{\rm e}$ in the lower panel
of Fig.\,\ref{ionfrac}. With the proper density parameter, we then calculated
the line luminosities of all [O{\sc i}] lines and derived the 
6300\,\AA/5577\,\AA~line ratio. This line ratio is plotted as a function of 
$q_{\rm e}$ in the upper panel of Fig.\,\ref{ionfrac}. In this plot we also 
included the observed line ratio, shown as the dotted line. 

Our calculations cover a temperature range between 7500\,K and 9000\,K. The 
curves in the upper panel of Fig.\,\ref{ionfrac} all show the same trend of a 
decrease in line ratio with increasing ionization fraction. This effect is 
clear, because the number of free electrons determines the collisional 
excitation rate of the levels. The more electrons, the higher the probability 
for collisional excitation even of the level with highest energy (i.e. level 5) 
from which the 5577\,\AA~line arises. The second obvious effect is the decrease 
in line ratio with increasing temperature, since the temperature determines the 
kinetic energy of the free electrons. For low temperatures, the energy of the 
free electrons is not high enough to effectively excite level 5, reducing the 
5577\,\AA~line emission and hence increasing the line ratio. 

For the three temperature values used to calculate the theoretical curves in 
Fig.\,\ref{ionfrac} we find agreement with the observed line ratio for the 
following parameter combinations: for 9000\,K we find an ionization fraction of
$10^{-5}$, and for 8000\,K we find an ionization fraction of $4\times 10^{-4}$.
For a temperature of 7500\,K, we cannot reproduce the observed line ratio.

Our calculations show that when accounting for the line luminosity and line 
ratio observed from R\,126, the ionization fraction is rather low for high 
temperatures, while an ionization fraction of up to 10\% might exist for a 
temperature as low as 7800\,K. This behaviour is only strictly valid for
a given observed line luminosity. In order to restrict the range of valid
or reasonable disk parameters, we take account of the fact that the
ionization fraction can only increase with increasing temperature, since 
collisional ionization will act more efficiently. This indicates that for the 
disk around R\,126 we can exclude high temperatures ($\ga 9000$\,K) as well as 
low temperatures ($\la 8000$\,K), and we conclude that the [O{\sc i}] line 
forming region in the disk of R\,126 has a temperature range of $\sim 8000
\ldots 8500$\,K with a corresponding ionization fraction range of $5\times
10^{-4} \ga q_{\rm e} \ga 5\times 10^{-5}$. This range in ionization fraction 
states that less than 0.05\% of hydrogen in the disk is ionized. We can 
therefore conclude that the disk around R\,126 is indeed neutral in hydrogen.

\begin{figure}[t!]
\resizebox{\hsize}{!}{\includegraphics{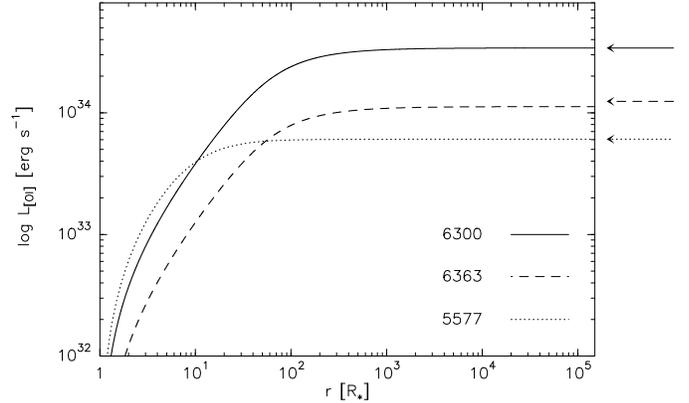}}
\caption{Calculated increase in [O{\sc i}] line luminosities from the hydrogen
neutral disk around R\,126. The observed, extinction corrected values are
indicated with arrows. The observations have been taken with an aperture radius
of 2\arcsec, corresponding to an outer edge of the observed wind of about
$1.5\times 10^{5}$\,R$_{\odot}$ at the distance of the LMC. This value is used
as the extent of the line luminosity integration, but the [O{\sc i}] line
luminosities clearly saturate already within about 500\,R$_{*}$.}
\label{oi_fit}
\end{figure}

Fig.\,\ref{oi_fit} shows the increase in [O{\sc i}] line luminosities 
calculated for a disk of 8000\,K and an ionization fraction of $4\times 
10^{-4}$. The lines saturate within a few hundred stellar radii. The 
observed values are indicated by the arrows to the right.

\subsection{Disk outflow velocity and mass loss rate}\label{mdot_const}

For a temperature of 8000\,K as derived in the previous section, we end up with 
a density parameter of $F_{\rm m, disk}/v_{\rm out}\simeq 
2.2\times 10^{-11}$\,g\,cm$^{-3}$
for the disk of R\,126 (see Fig.\,\ref{ionfrac}). To compare our results with 
those of Porter (\cite{John}), we need to convert this density parameter into 
a mass loss rate. For this, the outflow velocity, $v_{\rm 
out}$, of the disk is needed. For a pole-on seen disk with radially outflowing 
material, the maximum observable line-of-sight velocity, $v_{\rm los}$, is 
linked to the outflow velocity via $v_{\rm los} = v_{\rm out} \sin\alpha$ (see 
Fig.\,\ref{sketch}). The range in possible $v_{\rm los}$ values over the 
different disk outflow directions above and below the disk mid-plane leads to 
symmetrical doppler shifts of the line center, while thermal and turbulent 
velocities as well as the spectral resolution of the instrument result in a 
Gaussian profile of the line. Thus, optically thin lines formed in a pole-on 
seen outflowing disk should be symmetric with respect to their laboratory 
wavelength. The observed [O{\sc i}] lines are indeed symmetric 
(Fig.\,\ref{oi_r126}) and their profiles show a Gaussian shape with a FWHM of 
$\sim 9.2$\,km\,s$^{-1}$. 

\begin{figure}[t!]
\resizebox{\hsize}{!}{\includegraphics{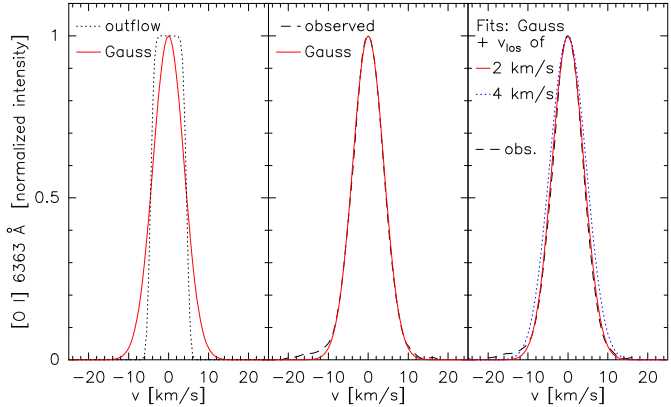}}
\caption{Normalized profiles of the [O{\sc i}] 6363\,\AA~line. {\it Left:}
Theoretical profiles for either a pure Gaussian or a pure outflowing disk
profile. {\it Middle:} Fit of a Gaussian profile with FWHM of 9.2\,km\,s$^{-1}$
resulting from FEROS' spectral resolution plus thermal motion of the gas. 
{\it Right:} Fit of the same Gaussian including an outflow component causing 
$v_{\rm los}$ of either 2\,km\,s$^{-1}$ or 4\,km\,s$^{-1}$. While the first one 
still gives a reasonable fit, the latter one clearly results in a too broad 
line profile.}
\label{profiles}
\end{figure}

To derive a possible outflow contribution from the observed lines, 
we first calculate the following test profiles: 
(i) a pure Gaussian line profile and (ii) a pure outflowing disk profile.
The results are shown in the left panel of Fig.\,\ref{profiles}. The calculated 
lines are normalized to their maximum value, and $v_{\rm gauss}$ and $v_{\rm 
out}$ are chosen such that they result in a FWHM value of 
9.2\,km\,s$^{-1}$. The pure outflowing disk (neglecting the influence of FEROS' 
spectral resolution) results in a flat-topped profile with steeply rising wings.
The narrow observed [O{\sc i}] lines show a better agreement with a 
Gaussian line shape, and the fact that the observed [O{\sc i}] lines have
widths that are only marginally broader than the velocity resolution of FEROS
indicates that the intrinsic FWHM of the lines can only be on the order of
2-3\,km\,s$^{-1}$. Such a low intrinsic FWHM can be ascribed to pure
thermal broadening. 

Since spectra taken with FEROS are not flux-calibrated, the reproduction of
the line profiles can only give qualitative results. We took exemplarily the 
line profile of the 6363\,\AA~line obtained with FEROS, corrected it for the LMC
radial velocity component of 260\,km\,s$^{-1}$, and normalized the line to its
maximum intensity. This normalized line profile is what we intend to fit.

We first used a pure Gaussian line profile with FWHM of 9.2\,km\,s$^{-1}$ 
(resulting from FEROS' spectral resolution plus thermal broadening of the line) 
which results in a good fit (mid panel of Fig.\,\ref{profiles}).  

\begin{figure}
\resizebox{\hsize}{!}{\includegraphics{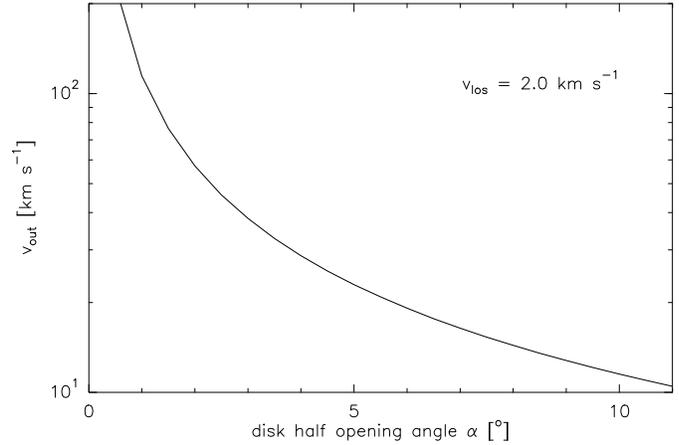}}
\caption{Disk outflow velocity as function of disk half opening angle for
$v_{\rm los} = 2$\,km\,s$^{-1}$ as derived from the [O{\sc i}] line profile 
fitting.}
\label{alpha_vout}
\end{figure}

Next we combined this Gaussian profile with an outflowing disk profile for
different values of $v_{\rm los}$. This was done to see for which value of
$v_{\rm los}$ the line profile starts to alter. We found that a possible 
$v_{\rm los}$ contribution must be smaller than $\sim 2$\,km\,s$^{-1}$ (right 
panel of Fig.\,\ref{profiles}) resulting in an outflow velocity of $v_{\rm 
out}\la 11.5$\,km\,s$^{-1}$ for $\alpha = 10\degr$. The value of $\alpha$ 
should be taken as an upper limit since the range of possible B[e] supergiant 
disk values has been estimated by Zickgraf (\cite{Zick92}) to be $\alpha = 
5-10\degr$. A thinner disk will result in a higher outflow velocity for the 
same value of $v_{\rm los}$ (see Fig.\,\ref{alpha_vout}). Nevertheless, even 
considering a thinner disk, the outflow velocity is considerably smaller than 
the values of 60-80\,km\,s$^{-1}$ usually ascribed to the disks of B[e] 
supergiants. In the case of R\,126, no good estimates for the real disk outflow 
velocity exist and only a detailed investigation of the line profiles from
emission lines that might be connected with the disk will help to clarify this
point.

An outflow velocity of 11.5\,km\,s$^{-1}$ as derived above together with the 
disk density parameter derived from the [O{\sc i}] line luminosity calculations 
and a disk opening angle of 20\degr~results in a total disk mass loss rate 
of $\dot{M}_{\rm disk} \simeq 2.5\times 10^{-5}$\,M$_{\odot}$yr$^{-1}$. This 
is about a factor of 10 higher than the value used by Porter\footnote{Note 
that the parameter $\dot{M}_{\rm d}$ used by Porter does not describe the {\it 
total} disk mass loss rate. Instead, his value of $\dot{M}_{\rm d} = 
10^{-5}$\,M$_{\odot}$yr$^{-1}$ needs to be multiplied by the solid angle of 
the disk and divided by $4\pi$. The total disk mass loss rate in his models is 
therefore $\dot{M}_{\rm disk}\simeq 2\times 10^{-6}$\,M$_{\odot}$yr$^{-1}$.}
(\cite{John}), and well in agreement with the postulated need for a higher disk 
density in order to account for the observed [O{\sc i}] line luminosities of 
R\,126. 

Based on recent observations with the {\it Spitzer Space Telescope} Infrared
Spectrograph, Kastner et al. (\cite{Kastner}) derived a total dust mass of the 
disk around R\,126 of $\sim 3\times 10^{-3}$\,M$_{\odot}$, and they claimed 
that the dust is contained between 120\,AU and 2500\,AU.
In order to compare this dust mass with our derived disk mass loss rate, we 
assume a gas to dust ratio of about 100 and convert the total mass into a disk 
mass loss rate for our outflowing disk model. We end up with a density 
parameter of $F_{\rm m, disk}/v_{\rm out}\simeq 3\times 10^{-10}$g\,cm$^{-3}$
and a disk mass loss rate of $\dot{M}_{\rm disk}\simeq 3.4\times 
10^{-4}$\,M$_{\odot}$yr$^{-1}$. This is about 170 times higher than the value
used by Porter (\cite{John}) and about 13 times higher than our value, 
and confirms that the disk around R\,126 must indeed be more massive than
previously thought.

In addition, the inner edge of the dusty disk of 120\,AU, which corresponds to 
about 360\,R$_*$, lies within our [O{\sc i}] saturation region. This 
confirms that our value for the disk mass loss rate is indeed a lower limit.

\section{The polar wind}\label{polar}
                                                                                
With our assumption of a hydrogen neutral disk right from the stellar surface, 
the free-free and free-bound emission is restricted to the ionized wind parts 
(see Fig.\,\ref{sketch}). At all latitudes above and below the equatorial disk 
we assume that the star has a normal B-type line-driven wind, which we refer to 
as the polar wind. Its radial density distribution, following from the equation 
of mass continuity, is
\begin{equation}
n_{\rm H}(r) = \frac{1}{\mu m_{\rm H}}\left(\frac{R_*}{r}\right)^{2}
\frac{F_{\rm m, pol}}{v(r)}\,, 
\label{masscont}
\end{equation}
where $F_{\rm m, pol}$ is the mass flux of the polar wind.
The velocity increase in line-driven winds can be approximated with a 
$\beta$-law with $\beta$ typically in the range of 0.8-1.0 for hot star winds 
(see e.g. Lamers \& Cassinelli \cite{LamCas}). We assume that the wind 
temperature is constant which allows for a simplified treatment of the 
intensity calculation. At every point along the impact parameter, $\zeta$ (see 
Fig.\,\ref{sketch}), the intensity is given by 
\begin{equation}
I_{\nu} = B_{\nu}(T)\left( 1 - e^{-\tau(\zeta)}\right)\,.
\end{equation}
The optical depth $\tau_{\nu}(\zeta) = \int \kappa_{\nu}(\zeta,s) ds$ is 
defined by the line-of-sight integral (with the line-of-sight perpendicular
to the disk mid-plane) over the absorption coefficient, $\kappa_{\nu}$, 
defined by
\begin{equation}
\kappa_{\nu}(\zeta,s) \simeq 3.692\times 10^{8}\,\frac{n_{e}^{2}(\zeta,s)}
{\nu^{3}T^{1/2}}\left(1-e^{-\frac{h\nu}{kT}}\right)\left(g_{\rm ff} + g_{\rm 
fb}\right)\,,
\end{equation}
and integration limits depending on the geometry of the system 
(Fig.\,\ref{sketch}). The parameters $g_{\rm ff}$ and $g_{\rm fb}$ are the
gaunt factors for free-free and free-bound processes, respectively, and have 
been calculated (e.g. by Kraus \cite{PhD}) over a large frequency range and 
for different temperatures. Due to the symmetry and the assumed constant wind 
temperature, the total observable flux density of the free-free 
and free-bound emission is given by
\begin{equation}
F_{\nu} = \frac{2\pi}{d^{2}}\int\limits_{0}^{\zeta_{\rm
max}} B_{\nu}(T)\left( 1 - e^{-\tau(\zeta)}\right) \zeta d\zeta
\end{equation}
where $\zeta_{\rm max}$ is the outer edge of the ionized wind and $d$ is the
distance to the object. Due to the pole-on view,
the dusty disk around R\,126 might cause absorption of the free-free
and free-bound emission from the rear parts of the wind (i.e. below the disk). 
However, as stated by Kastner et al. (\cite{Kastner}), the dust is expected to 
exist only at large distances, i.e. beyond 120\,AU. For the wind densities of
our interest, we found that the free-free and free-bound emission is generated 
much closer to the star, and any absorbing influence of the dust can be 
neglected.

\begin{figure}[t!]
\resizebox{\hsize}{!}{\includegraphics{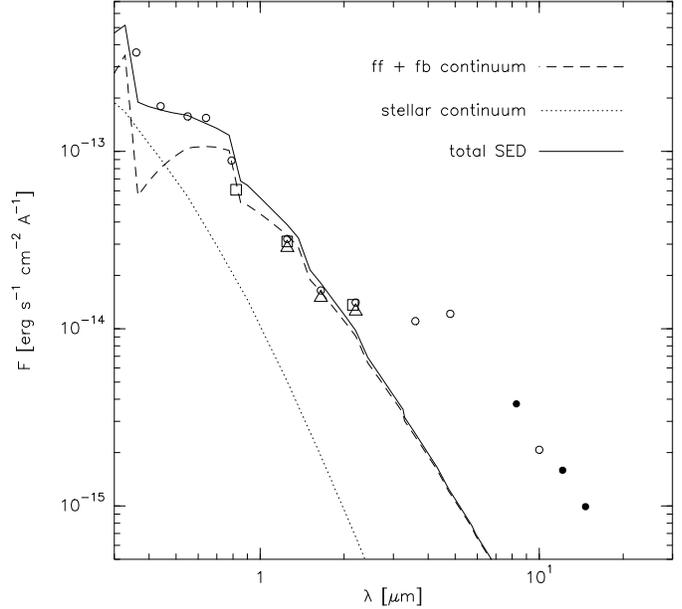}}
\caption{Calculated SED of R\,126 consisting of the stellar continuum and the
free-free and free-bound continuum from the polar wind. The theoretical spectrum
is reddened to the R\,126 value. Photometric data are from Zickgraf et al.
(\cite{Zickgraf}, open circles), and from the public databases DENIS (squares),
2MASS (triangles), and MSX (filled circles). Not included here is the modeling 
of the dust emission to account for the mid-IR excess.}
\label{SED}
\end{figure}

To calculate the electron density at every point in the wind, we need 
to know the polar mass loss rate and terminal velocity. Literature values for 
R\,126 are still not well constrained and vary over a huge range, depending on 
the geometry and the method used by individuals to derive them. As an example 
serves the terminal velocity for which Bjorkman (\cite{Bjorkman}) gives 
650\,km\,s$^{-1}$ while Zickgraf et al. (\cite{Zickgraf}) found 
1800\,km\,s$^{-1}$. Due to this uncertainty by almost a factor of 3 we 
calculate the free-free and free-bound emission by varying the density 
parameter, i.e. the ratio $F_{\rm m, pol}/v_{\infty}$, instead of the 
individual parameters $F_{\rm m, pol}$ and $v_{\infty}$.
The total continuum, including the stellar emission,
is then reddened with the extinction of R\,126, and compared with 
photometric data from the literature. A good fit to the SED in the optical 
and near-IR is achieved for $F_{\rm m, pol}/v_{\infty} \simeq 2.0\times 
10^{-12}$\,g\,cm$^{-3}$ and $\beta = 0.8$ (see Fig.\,\ref{SED}). 

\begin{table}[ht]
\caption{Model parameters for the polar and equatorial (i.e. the disk-forming)
winds. Please note that the disk parameters only hold for the [O{\sc i}] line
forming region.}
\label{param}
  \begin{tabular}{l|ccc}
         \hline
         \hline
          Wind & $T_{\rm e}$ & $F_{\rm m}/v_{\infty}$ & $q_{\rm e}$ \\
               & [K] & [g\,cm$^{-1}$] &  \\
         \hline
          Polar        & 10\,000 & $2.0\times 10^{-12}$ & 1.0 \\
          Disk-forming & 8\,000  & $2.2\times 10^{-11}$ & $4.0\times 10^{-4}$ \\
         \hline
  \end{tabular}
\end{table}

The density parameters for the equatorial and polar wind as summarized in 
Table\,\ref{param} result in a the density contrast on the order of $\rho_{\rm 
eq}/\rho_{\rm pole} \simeq 10$. This value turns out to be a {\it lower 
limit}, because all parameters are chosen such that any change
(like (i) a smaller disk opening angle, (ii) the existence of a small
ionized inner disk part, (iii) a vertical density distribution within the disk
due to a latitude dependence of the parameter $n_{\rm H}(R_{\rm in})$, (iv)
a cut-off in [O{\sc i}] line luminosities before saturation is reached) will
immediately request a higher disk density in order to reproduce the observed
[O{\sc i}] line luminosities.

With a disk density parameter of 10 times the polar one, but a disk ionization 
fraction of less than $10^{-3}$, the radial electron density distribution  
within the disk is at least a factor of 100 lower than the one in the polar 
wind. A contribution of the disk to the free-free and free-bound continuum is 
therefore indeed negligible, justifying our assumption that the free-free and 
free-bound emission is generated purely within the polar wind.

\section{Discussion}\label{discussion}

For our description of the disk model in Sect.\,\ref{geometry} we made two 
severe assumptions, namely we requested that
(i) the disk is neutral in hydrogen right from the stellar surface, i.e. 
$R_{\rm in} \simeq R_*$, and that
(ii) the disk has a constant outflow velocity. We will now discuss the validity 
of these assumptions.
 
{\bf A hydrogen neutral disk right from the stellar surface.}
We requested that the disk around the rather hot central star is predominantly 
neutral in hydrogen right from the stellar surface. Recent ionization structure 
calculations by Kraus \& Lamers (\cite{KL03}) in non-spherical winds of B[e] 
supergiants have shown that in the equatorial region hydrogen can indeed 
recombine close to the stellar surface, resulting in a geometrically thick 
hydrogen neutral disk. To quantify this, we refer to their model C which was 
calculated for a star with the same effective temperature as R\,126. Further 
parameters given in their Table\,1 were an electron temperature of 10\,000\,K, 
and an equatorial surface density of $3.3\times 10^{12}$\,cm$^{-3}$. With these 
parameters, Kraus \& Lamers (\cite{KL03}) found that hydrogen recombines in the
equatorial plane at a distance of $r \la 1.0004$\,R$_*$.

Our model calculations for the [O{\sc i}] line luminosities revealed a 
disk density parameter of about $2.2\times 10^{-11}$\,g\,cm$^{-3}$ (see 
Table\,\ref{param}) which results in an equatorial density on the stellar 
surface of $n_{\rm H}(R_*)\simeq 10^{13}$\,cm$^{-3}$. This density is about 3 
times higher than the one used by Kraus \& Lamers (\cite{KL03}). Consequently, 
recombination of the disk material will happen even closer to the stellar 
surface. It is therefore reasonable to adopt that the disk around the B[e] 
supergiant R\,126 is neutral in hydrogen right from the stellar surface.

{\bf The constant disk outflow velocity.} In order
to describe the disk density distribution with only one free parameter, i.e. 
the density parameter $n_{\rm H}(R_{\rm in}) \sim F_{\rm m, disk}/v_{\rm out} 
= {\rm const}$, we set the disk outflow velocity constant within the 
[O{\sc i}] line forming region. Fig.\,\ref{oi_fit} shows that the line 
luminosities of the [O{\sc i}] lines saturate within about 500\,$R_*$; the 
5577\,\AA~line saturates even within 50\,$R_*$. Let us assume that the 
disk-forming wind also has a velocity distribution according to the 
$\beta$--law. Such a velocity law then mainly influences the formation of the 
5577\,\AA~line. This line is created mainly in those wind parts in which
the wind is still accelerating, while the other two [O{\sc i}] lines are 
unaffected from a $\beta$--type velocity law. Their luminosities are produced 
at distances at which the velocity has reached already its terminal value. 

According to the equation of mass continuity, Eq.\,(\ref{masscont}), a lower 
velocity at a certain distance results in a higher wind density and therefore 
in a higher level population. Consequently, the luminosity of the 
5577\,\AA~line arising in the accelerated wind region is somewhat enhanced 
compared to the luminosity created in an outflow of constant (i.e. terminal) 
velocity. A higher line luminosity, however, results in a lower 
6300\,\AA/5577\,\AA~line ratio. Therefore, in order to fit the observations, 
a slightly different combination of disk ionization fraction and density 
parameter will be necessary.

We calculated the disk ionization fraction and density parameter for an 
outflowing disk with a temperature of 8000\,K and a velocity law with $\beta = 
0.8$. From our fitting procedure we found the following results: $q_{\rm e} 
\simeq 10^{-3}$ and $F_{\rm m, disk}/v_{\infty} \simeq 1.7\times 
10^{-11}$\,g\,cm$^{-3}$.
Even though the ionization fraction is a factor of 2.5 higher than in the case 
of the constant outflow velocity (see Table\,\ref{param}), the total amount
of ionized hydrogen in the disk is still only 0.1\%. In addition, the total
disk mass loss rate found with the $\beta$-type disk wind scenario is now
$\dot{M}_{\rm disk} \simeq 2\times 10^{-5}$\,M$_{\odot}$yr$^{-1}$. This
value is only slightly lower than the value derived with the constant outflow
velocity scenario in Sect.\,\ref{mdot_const}. Therefore, the conclusion of a 
hydrogen neutral disk with a density ratio of about a factor of 10 between 
equatorial and polar wind remains valid.

We want to emphasize that not much is known about the real velocity 
distribution of the outflowing disks around B[e] supergiants. Since a 
$\beta$--type velocity law does not drastically alter the results, the 
assumption of a constant outflow velocity as a first guess seems therefore 
to be a reasonable approach.

\section{Conclusions}\label{conclusions}
                                                                                
This paper is not aimed to present a detailed analysis of R\,126 and to derive
its wind and disk parameters with high accuracy, but to present a somewhat 
revised model for the outflowing disk scenario. We suggest that the disks 
around B[e] supergiants are neutral in hydrogen at, or close to the stellar 
surface. Indications for such a model come from observations as well as
from theory and are here briefly summarized:
\begin{itemize}
\item Former investigations of the disks around B[e] supergiants assumed 
that the observed near-IR excess is caused by free-free emission from the disk,
as it is the case for classical Be stars. A contribution from the polar wind
was thought to be neglibible. Fitting of the near-IR excess with free-free 
emission arising purely in the outflowing disk delivers a disk mass loss 
rate that defines an upper limit to the disk (dust) density. 
The resulting dust emission was however found to be too low to account 
for the observed IR excess (Porter \cite{John}).
\item Porter (\cite{John}) tried to solve this dust emission problem by 
allowing for a flatter disk density distribution. However, his modified disk 
model for the B[e] supergiant R\,126 is not able to reproduce the observed 
[O{\sc i}] line luminosities, underestimating them by about a factor of 50! To 
account for the [O{\sc i}] line luminosities a (much) higher disk density is 
needed.
\item Due to the nearly equal ionization potentials of O{\sc i} and H{\sc 
i}, the existence of strong [O{\sc i}] emission further requests that in the
line forming region hydrogen must be predominantly neutral. 
\item Recent ionization structure calculations in non-spherically symmetric 
winds of B[e] supergiants have revealed that the wind material indeed
recombines in the equatorial plane already at, or very close to the stellar 
surface, resulting in a hydrogen neutral equatorial disk or zone (see e.g.
Kraus \& Lamers, \cite{KL03}, \cite{KL06}; Kraus \cite{Kraus06}).
\end{itemize}
The most important consequence of such a hydrogen neutral disk is the fact
that the disk can no longer be considered as the formation location of the 
free-free emission causing the near-IR excess. This emission must then mainly 
be produced in the line-driven polar wind. Consequently, the fitting of the 
free-free emission does not result in an upper limit to the disk mass loss rate 
anymore. Instead, the disk mass loss rate, and hence the disk density, can be 
(much) higher, providing an ideal environment for the [O{\sc i}] line emission
as well as for efficient dust condensation to account for the observed IR 
excess.

We tested the scenario of a hydrogen neutral disk for the LMC B[e] 
supergiant R\,126 by modeling the line luminosities and line ratios of the
[O{\sc i}] emission lines resolved in our high-resolution optical spectra. 
The parameters derived for the disk and wind of R\,126 are the following:
\begin{itemize}
\item We found that the [O{\sc i}] 6300\,\AA/5577\,\AA~line ratio is
very sensitive to the ionization fraction in the disk. From fitting the observed
line ratio we can conclude that hydrogen in the disk is ionized by less than
0.1\%. This confirms that the disk is indeed predominantly neutral in hydrogen.
\item The disk mass loss rate of $\dot{M}_{\rm disk} \ga 2.5\times
10^{-5}$\,M$_{\odot}$yr$^{-1}$ found from our fitting is about a factor
of 10 higher than the value used by Porter (\cite{John}). This is in good 
agreement with the postulated need for a higher disk density in order to fit 
the [O{\sc i}] lines, and also in good agreement with the total dust 
mass of about $3\times 10^{-3}$\,M$_{\odot}$ as derived by Kastner et al.
(\cite{Kastner}). A disk with (much) higher density provides a much 
better environment for efficient dust condensation than the disk model used
by Porter (\cite{John}). 
\item The near-IR excess is fitted with free-free and free-bound emission
from the B-type line-driven polar wind. The resulting density contrast between 
equatorial and polar wind is on the order of 10. This value is found to be 
a lower limit.  
\end{itemize}
To summarize, based on a detailed investigation of the emerging [O{\sc i}] 
emission lines we found that the disk around the B[e] supergiant R\,126 is
neutral in hydrogen right from the stellar surface. Since all B[e] supergiants 
show strong [O{\sc i}] line emission, we postulate that the hydrogen neutral 
ouflowing disk scenario might also hold for the other members of the B[e] 
supergiant class.

                                                                                
\begin{acknowledgements}
                                                                                
We thank the anonymous referee for suggestions and critical comments 
that helped to improve the paper. 
M.K. acknowledges financial support from GA \v{C}R 205/04/1267.
M.B.F. is supported by \emph{CNPq} (Post-doc position - 150170/2004-1).
                                                                                
\end{acknowledgements}
                                                                                

\end{document}